\documentclass[12pt,a4paper]{article}
\usepackage{amsmath,amssymb}
\usepackage{exscale}
\usepackage{relsize}

\newtheorem{theor}{Theorem}

\def\endproof{\par\nobreak\hbox to \hsize{\hfil\vrule width 5pt height
5pt}\goodbreak\vskip 3pt}

\begin{document}

\title{On the Mathematics of Thermodynamics}
\author{J. B. Cooper\\ Johannes Kepler Universit\"at Linz
\and
T. Russell\\Santa Clara University }
\date{}
\maketitle

\tableofcontents
\newpage
\begin{abstract}

We show that the mathematical structure of Gibbsian thermodynamics 
flows from the following simple elements:
the state space of  a thermodynamical
substance is a measure space $\Omega$ together with two
orderings (corresponding to \lq \lq warmer than'' and \lq \lq adiabatically 
accessible from'') which satisfy certain plausible physical axioms and
an area condition which was introduced by Paul Samuelson.
We show how the basic identities of thermodynamics, in particular the 
Maxwell relations, follow and so the existence of energy, free energy, 
enthalpy and the Gibbs potential functions.  We also discuss some questions 
which we have
not found dealt with in the literature, such as the amount of information
required to reconstruct the equations of state of a substance and a systematic
approach to thermodynamical identities.  We illustrate the power of these
methods by giving in detail explicit computations for several real gases
which,
in the textbooks, are normally only
obtained for the simple case of an ideal gas. 
\end{abstract}

\thanks{This article is dedicated to the memory of Paul A. Samuelson,
who had the initial insight which led to our approach and who
accompanied its development with a barrage of new questions, ideas 
and encouragement.}

\section{Introduction}
The subject of thermodynamics is notoriously difficult for mathematicians.
V.I. Arnold [Ar] famously put it in a nutshell as follows:
 
\begin{quote}
Every mathematician knows that it is impossible to understand 
any elementary
course in thermodynamics.  
\end{quote}
He continues by explaining that

\begin{quote}
the reason is that [the] thermodynamics is based on a rather complicated 
mathematical theory, on [the] contact geometry.
\end{quote}

It is the purpose of this note to present an axiomatisation 
which is mathematically transparent, 
avoids anthropomorphisms, is elementary (no contact
geometry) and preserves all of the structure of the
classical theory.  It also allows us to carry out explicit computations
for the classical cases (ideal gas, van der Waals gas) in a simple and
general way which works for virtually any of the standard models for real gases. We   illustrate
this for an equation of state suggested by Feynman to allow for the
fact that, for real gases, the adiabatic index is far from being
constant and for a new model which combines the advantages of the
van der Waals and the Feynman gas.

We also discuss a theme which seems to us to be of eminent
practical significance but which we have never seen treated
in the literature---namely, how much information is required
to recreate the equations of state of a substance.
Thus we show that knowledge of all of the isotherms and just two
adiabats (dually two  isotherms and the adiabats) suffices.

Our system of axioms has two basic ingredients---firstly a measure space, i.e.,
a set with a $\sigma$-finite positive measure (which describes
the possible states of a thermodynamic system), provided with
two pre-orderings, \lq \lq warmer than'' and \lq \lq adiabatically accessible 
from''. 
We show that if these orderings satisfy several physically plausible
conditions, then they are induced by numerical functions---empirical
temperature and empirical entropy in the thermodynamical case.

We then show that  an area condition which was made explicit by the 
distinguished
economist P.A. Samuelson ensures the existence of essentially unique
absolute temperature and entropy.  These have the property that the
four Maxwell relations then hold.  If we then interpret these as
integrability conditions,  we can deduce the existence of the 
four energy type functions of thermodynamics.  This is in contrast
to many standard treatments where these relations are deduced
from the existence of the energy functions via the Schwarz Lemma.
In a final section, we present a unified and systematic approach
to the classical
thermodynamical identities between derived quantities
and discuss some known and new models of real gases in much
more detail than one finds in standard treatments.

\section{Samuelson's vision}

Samuelson noted that
 classical thermodynamics and 
economics  are related by a common search for an 
optimising basis for observed behaviour.
In thermodynamics the observed isotherms and adiabats are hypothesised
to be derived from the minimisation of a scalar quantity \lq \lq energy''.
In economics, the observed input demand functions are hypothesised to be
 derived from the maximisation of the scalar quantity \lq \lq profits''.  
Deriving a test for these hypotheses then becomes a common task 
of both disciplines.

Thus we can interpret results of Maxwell  as establishing the equivalence
of the existence of an energy 
function which is minimised with the fact that the isotherms and the adiabatics
fulfill a simple and natural geometric condition which we called in an earlier
paper the 
\lq \lq $S$-condition''---see below for a precise formulation of the latter.

Such a condition is certainly implicit in Maxwell's argument.  However, it 
is not stated explicitly.  Samuelson claims no priority for having noticed 
via this diagram that constrained optimisation implies this relationship 
but we have been unable to find another reference stating the equilibrium
 condition in such a geometrically simple way.  Thus although the above area 
condition appears {\it implicitly} in many areas, we have found no previous 
instance of its explicit formulation in the literature.

In [Co1]  
we obtained a number of equivalent formulations
of this condition, notably a rather lengthy partial differential equation
in two arbitrarily chosen functions which have the given curves as contours
(corresponding to empirical temperature and entropy 
in the thermodynamical situation).
We then showed that when this equation is satisfied  there
are canonical recalibrations of empirical temperature and entropy for which
the  Jacobian is identically \lq \lq 1'' and this implies the existence
of an energy function and the validity of the Maxwell relations
for the recalibrated quantities. Note that, as in Maxwell's original
treatment, we derive his relations from an area condition in $(p,V)$-space,
the existence of recalibrations with the  $J=1$ condition being a consequence
of the $S$ condition.  This paper can thus be viewed as an attempt to 
rehabilitate and perhaps clarify the Maxwell approach.

We shall use this theory to discuss existence
and uniqueness of solutions to the above partial differential equation.
In particular, we show that given one family of curves (say the isotherms)
and two other members of the other family, then this uniquely determines the
family of adiabats (for a more precise mathematical formulation, see below).

The Maxwell/Samuelson area condition thus establishes an important duality 
between isotherms and adiabats and, more generally, between any two 
suitable families of 
level curves which are derived by minimizing energy in two different
 constraint regimes. These facts can be used to show how to exploit 
this duality to derive some explicit formulae for the dual functions
(adiabats given isotherms, isotherms given adiabats) for two standard textbook
cases, 
the ideal gas and the van der Waals gas.

Recently there has been renewed interest in the derivation of the
Maxwell relations in thermodynamics from
the Jacobian identity 
$\dfrac {\partial (p,V)}{\partial (T,S)} = 1$ (see, for example, [Ri]).  
As is well known, this
identity means that the the corresponding map from the  $(T,S)$-plane
into the $(p,V)$-plane is area preserving, and so this approach links
thermodynamics to such areas as geometrical mechanics where  
area-preserving mappings play a central role.
Maxwell was well aware that his relations had a 
geometric foundation and indeed he used Euclidean geometry \lq \lq to get 
his four
identities in an amazingly obscure way'' [Am].
Perhaps because of its opacity, 
Maxwell's use of geometry to derive his relationships seems to have largely 
disappeared from the physics textbooks.
In the discipline of economics, however, exactly this obscure geometrical 
argument of Maxwell was the subject of the 
acceptance speech of  Paul 
Samuelson, the first American Nobel laureate in
Economics.
The Maxwell/Samuelson area condition thus establishes an important duality 
between isotherms and adiabats and, more generally, between any two 
families of 
level curves which are derived by minimizing energy in two different
 constraint regimes.

\section{Orderings and the axiomatics}
In this section we discuss briefly a topic which is probably the 
basic problem of the 
theory of measurement---when is a physical quantity
described (in a meaningful fashion) by a number?  If this can be done, then
the values of the quantity in question can
be compared (as in \lq \lq warmer than'', \lq\lq worth more than'' etc.)
and the basic question is when the converse holds , i.e., when is such an 
ordering induced by a numerical function (\lq \lq temperature'', 
\lq \lq price'')?
This has been examined and re-examined countless times (to our 
knowledge, the first rigorous formulation of a mathematical theorem of 
this sort
was due to Debreu [De] who gave a sufficient condition for a preordering to
be induced by a utility function).  We give a brief discussion, firstly for
the sake of completeness and secondly because we would like to emphasise
what we regard as the central point, namely that the real line has a simple
and elegant characterisation as an ordered space.  The latter fact is wellknown
(although perhaps not quite as wellknown as it should be)
but we have never seen it used on the problem we are now addressing.
The opening pages of Maxwell's treatise [Ma] give a lucid treatment of 
temperature
as an  ordering.
\subsection{Orderings and utility functions}
Suppose that we are given a set $\Omega$ and a surjective mapping $f$ from
it onto the real line.  (Since we are only concerned with the 
order-theoretic aspects of the latter,
 we can, at will, replace it by any
order-isomorphic set, for example 
an open  interval, in particular the half-line).
Then $f$ induces a  preordering  $\leq_f$ where we  define:
$x\leq_f y$  if and only if $f(x)\leq f(y)$.
This preordering has the following properties:
\begin{enumerate}
\item
it is total and  has no largest or smallest element;
\item
it is order complete;
\item
$\Omega$ contains a countable, order dense subset.
\end{enumerate}
For our purposes it will be convenient to restate these properties in terms
of the family of predecessor sets of elements.  Thus, we define, 
for each $\alpha \in \Omega$ the sets
$$A_\alpha=\{x\in \Omega : x\leq_f \alpha\}  , \qquad U_\alpha
=\{x\in \Omega : x<_f \alpha\},$$
which have the following properties:
\begin{enumerate}
\item
The 
$A_\alpha$ are distinct and totally ordered by inclusion;
\item
the countable subfamily  $\{A_q: q \in \text {\bf Q}\}$
is order dense;
\item
$\bigcup A_\alpha = \Omega$, $\bigcap A_\alpha = \emptyset$;
\item
The family of the $A_\alpha$ is closed under intersections and  
for each $\alpha$, $$A_\alpha=\bigcap \{A_\beta:\beta > \alpha\}.$$
\end{enumerate}
The family of the $U$'s satisfies the corresponding properties, except that
4. is replaced by 
\begin{itemize}
\item[4'.]
the family  is closed under unions and 
for each $\alpha$, $U_\alpha=\bigcup \{U_\beta:\beta < \alpha\}$.
\end{itemize}
Further, if  $\alpha < \beta <\gamma$, then

$$A_\alpha \subset U_\beta\subset A_\beta\subset U_\gamma\subset A_\gamma.$$
(Note that if we start with a family  $A_\alpha$ as above, and define
$U_\alpha=\bigcup \{A_\beta:\beta < \alpha\}$, then the above
conditions are fulfilled).

We remark at this point that in this paper the inclusion $\subset$
will always be
exclusive, i.e., $A \subset B$ implies that $A$ and $B$ are distinct. 

\noindent
We are interested in the following converse statements:

\begin{theor} 
Suppose that we are given a family
$\cal A$ of subsets of $\Omega$ which is totally ordered by inclusion,
is closed under arbitrary intersections and satisfies the properties
\begin{enumerate}
\item if  $A \in \cal A$, then $A=\bigcap\{B\in {\cal A}, A \subset B\}$; 
\item there is a countable subset ${\cal A}_0$ which is order dense in $\cal A$;
\item $\bigcap \cal A = \emptyset$ and $\bigcup {\cal A} = \Omega$.
\end{enumerate}
Then there is a surjective mapping  $f$ from $\Omega$ onto
{\bf R} such that  $\cal A$ is the family $$\{f\leq \alpha: 
\alpha \in \text{\bf R}\}.$$

\end{theor}
The simple proof of this result follows from the following 
standard order-theoretical
characterisation of the real line:

\begin{theor}
Suppose that we have a set  $A$ with a total ordering such that
\begin{enumerate}
\item
$A$
has neither a smallest nor a greatest element;
\item
$A$ has a countable, order-dense subset;
\item
$A$ is order complete.

\end{enumerate}
Then  $A$ is order-theoretically isomorphic to the real numbers.
\end{theor}
This in turn follows easily from the following characterisation of the 
rationals, which  is due to Cantor:
\begin{theor}
Suppose that we have a set  $A$ with a total ordering so that
\begin{enumerate}
\item
$A$
has neither a smallest nor a greatest element;
\item
$A$ is countable;

\item
$A$ is dense in itself (i.e., if $x<z$ in $A$, then there is a $y\in A$
with $x<y<z$).
\end{enumerate}
Then  $A$ is order-theoretically isomorphic to the rational numbers.
\end{theor}

The idea behind the proof of theorem 1 is now simple.  
We introduce the following
equivalence relationship on  $\Omega$:  $x\sim y$ if and only if
$x$ and $y$ are in exactly the same sets of $A\in \cal A$, i.e., 
for each $A$, $x\in A$ if and only if $y\in A$.
Then the quotient space $\Omega|_\sim$ has a natural order structure
which satisfies the properties which characterise the real line.
The required mapping  $f$ is then the canonical one from $\Omega$
onto the quotient space. 

An important point is that the  $f$ in theorem 1 is not uniquely determined.
We can replace
it by any $F= \phi \circ f$ where $\phi$ is an arbitrary order isomorphism of 
the line.  We call such an $F$ a {\it recalibration} of $f$.
In the general situation there is no canonical choice of $F$.
This  will be crucial in the following.

There are two refinements of theorem 1 which will be of particular
interest to us.  Firstly, if $\Omega$
is provided with a suitable $\sigma$ algebra and each $A\in\cal A$ is 
measurable (i.e., a member of the algebra---we then say that the 
ordering is {\it measurable}),  $f$ will be measurable. 
Secondly, if $\Omega$
is a topological space and each $A \in \cal A$ is closed and, further,
for each $A \in \cal A$, $U=\bigcup \{B\in {\cal A}: B \subset A\}$
is open, then $f$ is continuous.  There are corresponding 
conditions which ensure
semi-continuity.

\subsection{The axiomatics}

We are now in a position to state the four axioms which describe
the mathematical structure of a thermodynamical theory:

\begin{enumerate}
\item The states of a thermodynamical system are specified by the points
of a set $\Omega$ with a positive $\sigma$-finite measure $\mu$.
\item  $\Omega$ is provided with two  families $\cal A_{\text{temp}}$
and $\cal A_{\text{ent}}$ of measurable subsets
which satisfy the conditions of theorem 1 (and so the preorderings 
are induced by
numerical functions which we denote by $t$ and $s$).
\item 
for each $A_{-1}\subset A_0 \subset A_1$ in
$\cal A_{\text{temp}}$
and $B_{-1}\subset B_0 \subset B_1$ in $\cal A_{\text{ent}}$ we have
\begin{eqnarray*}
&\mu((A_1\setminus A_0)\cap(B_1\setminus B_0))
\mu((A_0\setminus A_{-1})\cap(B_0\setminus B_{-1}))\\
&\qquad \qquad \qquad =
\mu((A_1\setminus A_0)\cap(B_0\setminus B_{-1}))
\mu((A_0\setminus A_{-1})\cap(B_1\setminus B_0)).\end{eqnarray*}
\item
Condition  3. means that if  $t$ and  $s$ are the
(measurable) functions which induce these orderings,
then 
the image measure of  $\mu$ in $\text{\bf R}^2$ under the mapping
$\omega\mapsto(t(\omega),s(\omega))$ splits multiplicatively.
We further assume that this measure is equivalent to Lebesgue measure on the
plane (equivalent in the sense of being mutually absolutely continuous).
\end{enumerate}

It follows from this that for each
$A \in \cal A_{\text{temp}}$, and for each $A \in \cal A_{\text{ent}}$, 
$\mu (A \setminus \bigcup \{B\in {\cal A}: B \subset A\})=0$,
and that for each pair $A \subset A_1$ from $\cal A$ and 
$B \subset B_1$ from $\cal B$ we have
$$\mu((A_1\setminus A)\cap(B_1\setminus B))$$ is strictly positive and finite.

We  remark that these axioms are physically natural and  plausible.
(In classical thermodynamics, $\Omega$ is $(p,V)$-space and
the measure is interpreted as mechanical work).
Since this article is, despite its title, one in mathematics rather than
in physics, we will not go into this in detail; but we emphasise that the
area condition, in particular, is not a {\it deus ex machina}
inserted to save the day but has a natural physical justification. 
A further point, which is important more for philosophical reasons,
is that they do not explicitly refer to the real numbers (compare
the axioms for Euclidian geometry, particularly in the form as perfected by 
Hilbert).
 
 \subsection{The existence of 
absolute temperature and entropy}
If we only assume 1. of the above axioms,  
$s$ and $t$ are, as is the function $f$ in the general result,
not uniquely determined.  
A crucial point of our treatment is that in the presence of conditions 
3. and 4. above,
then there are (essentially) unique such choices, which we call
the {\it canonical recalibrations}.
For the condition 4. means that the image measure has the form 
$a(u)b(v)\,du\,dv$
where we denote the coordinates in $\bf R^2$ by $(u,v)$
and $a$ and $b$ are locally Lebesgue-integrable functions whose 
reciprocals are also locally
integrable.  Now if we replace the two functions $t$ and $s$ by the recalibrations
$T=\phi \circ t$ and $S=\psi \circ s$,  where $\phi$ is a primitive of 
$\dfrac 1a$
and $\psi$
of $\dfrac 1b$,
then we obtain the following result:
\begin{theor}
Suppose that the above axioms are satisfied.  Then we can choose the functions
$T$ and $S$ so that the mapping $\omega \mapsto (T(\omega),S(\omega))$
is area-preserving.  
\end{theor}

This choice is 
unique up to suitable affine transformations (loosely speaking,
we can choose the zero   point and a change of scale---c.f. the 
difference between the Celsius and Fahrenheit systems).

In thermodynamics, these canonical calibrations 
are called {\it absolute temperature} and {\it entropy}
(as opposed to empirical temperature and entropy). 

We now proceed to  show that these axioms imply the usual 
contents of elementary treatments of thermodynamics.

As we shall see shortly, this choice of calibration is crucial since
the fact that the area condition holds is equivalent to each (and hence all)
of the four Maxwell relations.  Since the latter can be interpreted as
integrability conditions, they ensure the existence of the four energy
type functions of thermodynamics.  (Once again, these are purely mathematical
facts, but the underlying motivation from thermodynamics is the principle of
Joule-Maxwell on the mechanical equivalence of heat).
\section{Samuelson configurations}
\subsection{The area condition}

We emphasise at this point that up till now smoothness (except in the
very mild form of measurability) has   played no part in our
considerations, neither in the formulation of the axioms nor in the
derivation of the canonical recalibrations.  This is as it should be, for 
philosophical
reasons but also because the presence of phase transitions makes it
clear that in real substances we can and should expect the isotherms and
adiabats to have \lq \lq corners''.
We now turn to the case where they {\it are} smooth, i.e., 
we suppose that $\Omega$ is the plane $\text{\bf R}^2$
or some suitable subspace (in thermodynamics usually the  positive quadrant)
and that the functions which induce the ordering are smooth (in the sense
of being infinitely differentiable) as are 
their level curves.  It is convenient
to use the mathematically neutral notation $x$ and $y$ for the coordinates
 in the plane which we shall initially regard as the independent variables
and $u$ and $v$ for
the two (potential) 
functions.
As shown in [Co1],  the area condition is equivalent to the
fact that the functions $u$ and $v$ satisfy
a certain non-linear partial differential equation of the third 
order which is displayed explicitly there.  Familes of level curves 
satisfying this partial differential 
equation are thus of great interest in the study of optimizing systems.  

In the following, we concentrate
on the two foliations consisting of the level curves of $u$ and $v$, i.e., 
the isotherms and adiabatics in the thermodynamical context. 
The area condition can then be formulated as follows.
The plane (or a suitable part thereof) is foliated by two families of curves--- 
the level curves of two potential functions $u$ and 
$v$.  We assume that these are transversal at each point, i.e., the Jacobian 
$J = (u_x v_y - u_y v_x)$ never vanishes.  Since the regions we 
consider are connected (in the topological sense),  
$J$ cannot change sign.  
Hence there is no essential loss of generality if we assume that it is 
always strictly positive.  

Locally the families of curves form a network 
which is topologically equivalent to the standard network of the plane induced
by the parallels to the $x$ and $y$ axes (i.e., the case where $u (x,y) = x$ 
and $v(x,y) = y$).

We say that the foliations   satisfy {\bf condition $S$} (or that $v$
 is $S$-transversal to $u$ or that the $v$-curves are $S$-transversal 
to the $u$ curves)
if the following holds: for any choice of values $c_{-1} < c_0 < c_1$, 
and $d_{-1} < d_0 < d_1$ respectively, we have

$$\mbox{area $A$/area $B$ = area $C$/area $D$}$$
where

\begin{eqnarray*}
A& =&\{ (x,y) : c_{-1} < u(x,y) < c_0, d_0 < v(x,y) < d_1\},\\
B &=&\{ (x,y) : c_{0} < u(x,y) < c_1, d_0 < v(x,y) < d_1\},\\
C& =&\{ (x,y) : c_{-1} < u(x,y) < c_0, d_{-1} < v(x,y) < d_0\},\\
D& =&\{ (x,y) : c_{0} < u(x,y) < c_1, d_{-1} < v(x,y) < d_0\}.
\end{eqnarray*}

In order to avoid topological problems, we assume that the values of the $c$'s
and $d$'s are sufficiently close for the above condition on the network 
to be satisfied.  This means that the condition we are considering is a 
local one, as it should be if it is to be equivalent to a partial 
differential equation.  However, it is easy to obtain a global form 
from the local one.

We refer to the configuration consisting of two foliations which 
are $S$-transversal as a {\it Samuelson configuration}.
The $S$-configuration consisting of the adiabats and isotherms of the
ideal gas is one of the most iconic
images of modern science.

  The precise relation of our result to this question will be made more 
explicit below.   For obvious reasons, we discuss the
case of the isotherms for an ideal gas and for a van der Waals gas in some 
detail.
In particular, we show that if a function $v$ is $S$-transversal to the 
function $u = xy$ (i.e., the potential defining the isotherms of an ideal gas) 
and {\it two} of $v$'s level curves have the form 
$x y^\gamma = \text {constant}$
for the same $\gamma$, then all of them have this form, i.e., the adiabatics 
are precisely those for the ideal gas with exponent $\gamma$.

\subsection{Thermodynamical notation}
Although this article is one on mathematics and not on physics,
its main motivation comes, of course, from thermodynamics.  
For this reason,  we recall the standard notation and 
concepts from classical thermodynamics for the reader's convenience.
The coordinates $x$ and $y$ in the neutral notation correspond to $p$
and $V$ in Gibbsian thermodynamics.  The choice of the latter
as independent variables is natural since these two quantities can be 
directly measured.
Also the natural meaure on this space (mathematically speaking, two-dimensional
Lebesgue
measure) has a natural physical interpretation (mechanical work).

Our starting point is the situation where we are 
given the temperature  $T$ and the entropy $s$ as functions of
the pressure $p$ and the volume $V$. 
We use the lower case $s$ to indicate that this is empirical
entropy.  Absolute entropy will be denoted by $S$.
(The standard models do not require a recalibration of temperature
so that there will be no need at this point to 
distinguish beween lower and upper
case $T$---however, we consider below an interesting model
due to Feynman where we {\it shall} require such a recalibration).
We use the following dictionary to jump between the purely 
mathematical notation and the thermodynamical one: 
$u$ corresponds to $T$, $v$ to $s$, $p$ to $x$ and $V$ to $y$.  For example,
 the
thermodynamical equations

$$T = p V, \quad s = pV^\gamma $$

\noindent of the ideal gas corresponds to

$$u(x,y) = x y, \quad v(x,y) = x y^\gamma.$$ 
For reasons which will be clear shortly, we use
the recalibrated form  
$$u(x,y) = x y,\quad  v(x,y) = \frac 1 {\gamma -1}\left (\ln x +
\gamma  \ln y \right )$$
of these equations.

It is a consequence of the Maxwell relations that one can define the 
following four energy type functions (whose definitions we repeat for the 
readers' orientation.  They can be found in any textbook on 
thermodynamics, e.g. [La]). 
Firstly, the {\bf energy $E = E(S,V)$} is a function of entropy and volume.  
From this 
one derives the quantities $T$ (temperature) and $p$ (pressure) by the 
equations

$$T = \frac{ \partial E(S,V)}{\partial S}, \quad p = -
\frac{ \partial E(S,V)}{\partial V}.$$
Analogously, one has the {\bf enthalpy} $H = H(p,S)$, from which one 
derives the quantities
$$T = \frac{ \partial H(p,S)}{\partial S}, \quad V = 
\frac{ \partial H(p,S)}{\partial p};$$
the {\bf free energy}  $F = F(T,V)$, from which one gets
$$S = -  \frac{ \partial F(T,V)}{\partial T}, \quad p = 
- \frac{ \partial F(T,V))}{\partial V};$$
and the {\bf free enthalpy}  $G = G(p,T)$, which gives
$$S = -  \frac{ \partial G(p,T)}{\partial T}, \quad V = 
 \frac{ \partial G(p,T))}{\partial p}.$$
In the German-language literature, e.g. [La],
one employs $\Phi$ for $G$ and $W$ for $H$.

We emphasise that in our treatment the logical development is reversed---the 
existence of such functions is a consequence of our axiom system, since
it follows from the 
Maxwell relations which in 
turn are equivalent to the validity of the $S_1$-condition (see below).


\subsection{Canonical recalibrations}
We saw above that if the level curves of 
$u$ and $v$
satisfy the $S$ condition, then we can find (essentially unique) recalibrations
$U = \phi \circ u$ and $V = \psi \circ v$ (where $\phi$ and $\psi$ are 
diffeomorphisms between, say, intervals of the real line), so that
the Jacobian is identically one.  If we assume that $u$ and $v$ are so 
calibrated, then this  means that the diffeomorphism
$(x,y) \mapsto (u(x,y),v(x,y))$ of the plane (or a suitable subset thereof)
is area preserving.  In this case we say that the functions $u$ and $v$ 
satisfy the {\bf  $S_1$-condition}, or that $v$ is {\bf $S_1$-transversal} to 
$u$.
(The recalibration for the ideal gas which was used above arose in this way).

We now come to the crucial point in our argument.
If we write the basic equations
$u=f(x,y)$, $v=g(x,y)$ in differential form, i.e., as
$$du=f_1\, dx+f_2\, dy,\quad  dv=g_1\, dx+g_2\, dy$$
($f_1$, $f_2$ are the partials with respect to $x$, $y$ etc.),
we can solve for  $du$ and $dx$, say,
to get
$$du=\frac{f_1}{g_1}\, dv-\frac J{g_1}\, dy, \quad dx=\frac
1{g_1}\, dv-\frac{g_2}{g_1}\, dy,$$ where $J$ is the Jacobi-determinant
$f_1g_2-f_2g_1$,
and so we see that the condition $J=1$ is equivalent to the Maxwell
relation $\dfrac{\partial u}{\partial y}\Big |_v=-\dfrac{\partial x}
{\partial v}\Big |_y$
which is an integrability condition and ensures the existence of a 
function $h$ of 
the two variables $y$ and $v$ such that
$u$ and $v$ are the solutions of the equations
$$x - f(y,v) = 0,\quad u - g(y,v) = 0, $$
where $f(y,v) = - \dfrac {\partial h}{\partial y}$ and
$g(y,v) =  \dfrac {\partial h}{\partial v}$.
(We are using the standard conventions employed in thermodynamics---thus
$\dfrac{\partial u}{\partial y}\Big |_v$ denotes the partial derivative
of $u$, regarded as a function of $y$ and $v$, with respect to $y$).

The proof of this result employs the inverse function theorem
and so the precise statement is local.  The same remark applies to many of 
the following enunciations.  

We shall call such a function $h$
a {\it geometric
energy function} since in certain situations where the foliations arise
as the level curves of suitable physical quantities it corresponds to
the energy of a system.  However, in such situations, the
energy function satisfies some structural properties (monotonicity,
convexity) which have natural physical interpretations, and these are
of no direct relevance in our considerations below.  In a similar manner,
we will talk of geometrical adiabatics associated 
with families of isotherms, or geometrical isotherms associated
 with families of adiabatics respectively.

The existence of the above energy functions is one of those facts
which have been discovered and rediscovered time and again in the history
of mathematics.  We have traced it as far back as to Gau\ss{} [Ga] 
who used it to
describe all equivalent projections (in the sense of mathematical cartography)
and it appears in contact geometry (under the name of a generating function).
Of course, as remarked above, it has long been used in thermodynamics.

It follows from the above observation that we have 
a remarkable symmetry (corresponding to the Maxwell relations in 
thermodynamics).  If we start with a given energy function, we can as above
calculate $u$ and $v$ as functions of $x$ and $y$.  The energy function 
arises from the process of replacing  $x$ and $y$ as independent 
variables  by $y$ and $v$.  There are four such possibilities (the 
interesting ones for us are those with  $y$ and $v$,
$x$ and $v$, $y$ and $u$ and  $x$ and $u$ 
respectively as independent variables), 
each of which is associated 
with an
\lq \lq energy function'' (as we noted above,
in thermodynamics they are called energy, free
energy, enthalpy and free enthalpy respectively).  Hence any one
 such  function automatically defines three others. (In fact, the situation 
is more complicated than described here.  This is due to the fact that 
we are relying on global solvability of the corresponding non-linear 
equations.  The general results we use employ the inverse function theorem 
and so only guarantee local solubility. In many concrete situations 
which we compute, we do, of course,  have global invertibility and hence
 the kind of 
symmetry evoked here).  

In the case of an ideal gas, the permutations of the various variables
can be computed by hand and are valid globally---we include the formulae 
below 
for completeness.  We have also added
a more general case since it displays
the fact that the familiar presence of the logarithm in the expression for 
the entropy of an ideal gas is in a certain sense unique to this case.
Already the van der Waals gas offers difficulties here and
we shall shortly develop an alternative method of computing these energy 
functions which is often more practical and doesn't require us to 
compute these permutations.

\section{Thermodynamical identities--- an 
{\it anthologie raisonn\'ee}}
\subsection{The basic machinery}

We suppose that $u$ and $v$ are given as functions $f$ and $g$ of $x$
and $y$. Thus $u=f(x,y)$, $v=g(x,y)$ and, when $J=1$, simple manipulations
with differential forms proved the basic identities:
$$\begin{array}{lclclcclclcl} 
du&=&f_1\, dx&+&f_2\, dy,  &&dv&=&g_1\, dx&+&g_2\, dy\\
&&&&&&&&&&\\
dx&=&g_2\, du&-&f_2\, dv, &&dy&=&-g_1\, du &+&f_1\, dv\\
&&&&&&&&&&\\
du&=&\frac{f_2}{g_2}\, dv&+&\fbox{\text{ $\frac 1{g_2}$}}\, dx, 
&&dy&=&\fbox{\text{$ \frac
1{g_2}$}}\, dv&-&\frac{g_1}{g_2}\, dx\\
&&&&&&&&&&\\
du&=&\frac{f_2}{g_1}\, dv&-&\fbox{\text{$ \frac 1{g_1}$}}\, dy, 
&&dx&=&\fbox{\text{$ \frac 1{g_1}$}}\, dv&-&\frac{g_2}{g_1}\, dy\\
&&&&&&&&&&\\
dv&=&\frac{g_1}{f_1}\, du&+&\fbox{\text{$ \frac 1{f_1}$}}\, dy, 
&&dx&=&\fbox{\text{$ \frac1{f_1}$}}\, du&
-&\frac{f_2}{f_1}\, dy\\
&&&&&&&&&&\\
dv&=&\frac{g_2}{f_2}\, du&-&\fbox{\text{$ \frac 1{f_2}$}}\, dx, 
&&dy&=&\fbox{\text{$ \frac
1{f_2}$}}\, du&-&\frac{f_1}{f_2}\, dx\end{array}$$ where  we have 
highlighted the expressions which correspond to the 
Maxwell relations.

\noindent In thermodynamic notation these are
$$\begin{array}{lclcclcl}
dT&=&f_1\, dp+f_2\, dV,&&dS&=&g_1\, dp+g_2\, dV\\
dp&=&g_2\, dT-g_2\, dS,&&dV&=&-g_1\, dT+f_1\, dV\\
dT&=&\dfrac{f_2}{g_2}\, dS+\dfrac 1{g_2}\, dp,
                        &&dV&=&\dfrac 1{g_2}\,
dS-\dfrac{g_1}{g_2}\, dp\\
dT&=&\dfrac{f_2}{g_1}\, dS-\dfrac 1{g_1}\, dV,&&dp&=&\dfrac 1{g_1}\,
dS-\dfrac{g_2}{g_1}\, dV\\
dS&=&\dfrac{g_1}{f_1}\, dT+\dfrac 1{f_1}\, dV,&&dp&=&\dfrac 1{f_1}\,
dT-\dfrac{f_1}{g_2}\, dV\\
dS&=&\dfrac{g_2}{f_2}\, dT-\dfrac 1{f_2}\, dp,&&dV&=&\dfrac 1{f_2}\, 
dT-\dfrac{f_1}{f_2}\, dp
\end{array}$$
where we are using the key: $u\leftrightarrow T$, $v\leftrightarrow
S$, $x\leftrightarrow p$, $y\leftrightarrow V$ introduced above. 

\noindent
In order to isolate the underlying patterns, we now use a numerical
code. Thus
$$\begin{array}{ccccc}
u&\rightarrow&3\leftarrow&T\\
v&\rightarrow&4\leftarrow&S\\
x&\rightarrow&1\leftarrow&p\\
y&\rightarrow&2\leftarrow&V.\end{array}$$

\noindent
Partial derivatives will be denoted by triples in  brackets. $(3,1,2)$, 
for example, denotes $\frac{\partial u}{\partial x}|_y$ 
in the neutral notation, $\frac{\partial T}{\partial p}|_V$
in the thermodynamical one.
In general, $(i,j,k)$ denotes the partial 
derivative of variable $i$, regarded as
a function of the $j$-th and $k$-th variable, with respect to the $j$-th
variable. 

By reading off from the above list, we can express each partial derivative 
of the form $(i,j,k)$ in
terms of $f_1,f_2,g_1,g_2$ as follows:
$$
(3,1,2)=f_1,\quad
(3,2,1)=f_2,\quad
(4,1,2)=g_1,\quad
(4,2,1)=g_2;$$

$$
(1,3,4)=g_2,\quad
(2,3,4)=-g_1,\quad
(1,4,3)=-f_2,\quad
(2,4,3)=f_1;$$

$$
(3,4,1)=\frac{f_2}{g_2},\quad
(2,4,1)=\frac 1{g_2},\quad
(3,1,4)=\frac 1{g_2},\quad
(2,1,4)=-\frac{g_1}{g_2};$$

$$
(4,3,1)=\frac{g_2}{f_2},\quad
(2,3,1)=\frac 1{f_2},\quad
(4,1,3)=\frac 1{f_2},\quad
(2,1,3)=-\frac{f_1}{f_2};$$

$$
(3,4,2)=\frac{f_1}{g_1},\quad
(1,4,2)=\frac 1{g_1},\quad
(3,2,4)=-\frac 1{g_1},\quad
(1,2,4)=-\frac{g_2}{g_1};$$

$$
(4,3,2)=\frac{g_1}{f_1},\quad
(1,3,2)=\frac 1{f_1},\quad
(4,2,3)=\frac 1{f_1},\quad
(1,2,3)=-\frac{f_2}{f_1}.$$

\noindent 
We can then express any derivative $(a, b, c)$ in terms of ones of the form
$(d, 1, 2)$ or $(e, 2, 1 )$.  Thus the four derivatives with $x$ and $v$ as 
independent
variables are as follows:

$$\begin{array}{rcr}
(3,4,1)&=&\dfrac{(3,2,1)}{(4,2,1)}\\
(2,4,1)&=&\dfrac 1{(4,2,1)}\\
(3,1,4)&=&\dfrac 1{(4,2,1)}\\
(2,1,4)&=&-\dfrac{(4,1,2)}{(4,2,1)}\end{array}$$

Then, as above, we can introduce four energy functions
$E^{13}$, $E^{14}$, $E^{24}$, $E^{14}$ such that $dE^{24}=u\, dv-x\, dy$, 
$dE^{14}=u\,dv+y\,dx$, $dE^{13}=-v\, du+y\, dx$, $dE^{23}=-v\, du-x\, dy$ 
(the superfixes correspond to the independent variables---thus for 
$E^{13}$ these are  $x$ and $u$ , i.e., $1$ and $3$).

We will discuss these in more detail below where the rationale of our notation 
will be explained. In terms of the classical 
notation:
$$\begin{array}{lcrcrl}
dE&=&T\, dS&-&p\, dV &\mbox{(energy)}\\
dF&=&-S\, dT&-&p\, dV &\mbox{(free energy)}\\
dG&=&-S\, dT&+&V\, dp &\mbox{(Gibbs potential)}\\
dH&=&T\, dS&+&V\, dp &\mbox{(enthalpy),}\end{array}$$
i.e., $E^{13}=G$, $E^{23}=F$, $E^{14}=H$ and $E^{24}=E$.

If we arrange the energy functions in lexicographic order , i.e., 
as $E^{13}$, $E^{14}$, $E^{23}$, $E^{24}$ and denote
them by $5$, $6$, $7$ and $8$ in this order, 
then we can incorporate them into our system.
For it follows from the definitions and simple substitutions that
\begin{eqnarray*}
dE^{13}&=& (y-v f_1)dx - f_2 v dy\\ 
dE^{14}&=&(u f_1+y) dx + u f_2 dy\\
dE^{23}&=&-v f_1 dx +(x-vf_2) dy\\
dE^{24}&=&uf_1 dx+(uf_2-x) dy
\end{eqnarray*}
and so
$$\begin{array}{cclcccl}
(5,1,2)& =& y-gf_1, && (5,2,1)&=& -gf_2\\
(6,1,2)&=&y+fg_1,&& (6,2,1)&=&fg_2\\
(7,1,2)&=&-gf_1,&& (7,2,1)&=&-x-gf_2\\
(8,1,2)&=&fg_1,&& (8,2,1)&=&-x+fg_2.
\end{array}$$

One of the potentially irritating features of the thermodynamical identities is
that many are related by a simple swapping of the variables while this
is accompanied by changes of sign which seem at first sight 
to be random.
The simplest example is displayed by the four Maxwell relations.
We can systemise such computations by 
introducing the symbol $[a,b;c,d]$ for the Jacobi determinant of the mapping
$(c,d)\mapsto (a,b)$, i.e.,
 $$[a, b;c, d] = (a, c, d)(b, d, c) - (a, d, c)(b, c, d).$$
The determinant then takes care of the sign.  

For example $[3,4;1,2]$ is the Jacobian
$\dfrac{\partial(u,v)}{\partial(x,y)}$ and  is therefore
$1$
(which here denotes the number $1$),
$[3, 2;4, 1]$ is $\dfrac{\partial(u,y)}{\partial(v,x)}$
and therefore $=-\dfrac{f_1}{g_2}=-\dfrac{(3, 1, 2)}{(4, 2, 1)}$.
\noindent
Note that there are $1,680$ such Jacobians.  However, lest the reader despair,
we then  have the following simple rules for manipulating these expressions
which allow us to express them all in terms of our primitive quantities ($f$
and $g$ together with their partials and, of course, $x$ and $y$).

$$[a, b;c, d]  = -[b, a;c, d] =-[a, b;d, c]$$
$$[c, d;a, b] = \dfrac1{[a, b;c, d]}$$
$$(a,b,c)=[a,c;b,c]$$
Further useful rules for computation are
$$ [a,b;c,d]=\dfrac{[a,b;e,f]}{[c,d;e,f]},$$
in particular,
$$ [a,b;c,d]=\dfrac{[a,b;e,b]}{[c,d;e,b]}$$
and
$$ [a,b;c,d]=\dfrac{[a,b;1,b]}{[c,d;1,b]}.$$

Using these rules, we can 
compute any of the 336 expressions of the form $(i,j,k)$
(for $i$, $j$ and $k$ running from $1$ to $8$)
by routine computations.
For example, if we wish to compute $(8,3,5)$ then we proceed as follows:
Firstly, 
$$ (8,3,5)=[8,5;3,5]=\dfrac{[8,5;1,2]}{[3,5;1,2]}$$ 
Now 
$[8,5;1,2]=(8,1,2)(5,2,1)-(8,2,1)(5,1,2)$ 
and 
$[3,5;1,2]=(3,1,2)(5,2,1)-(3,2,1)(5,1,2)$.

Hence, finally
$$(8,3,5)=\frac{(8,1,2)(5,2,1)-(8,2,1)(5,1,2)}{(3,1,2)(5,2,1)-(3,2,1)(5,1,2)} $$
and so can be expressed in terms of $f$  and $g$, together with their partials.

\noindent
As a simple example, we can  compute again the basic formulae
$$(4, 3, 1) = \dfrac{g_2}{f_2}, \quad (4, 1, 3) = -\dfrac1{f_2}, \quad
(2, 3, 1)=\dfrac1{f_2},\quad(2, 1, 3)=
\dfrac 1{f_2}.$$
For example $$(4, 3, 1)=[4, 1;3, 1]=\dfrac{[4, 1;1,2]}{[3, 1;1, 2]}=
\dfrac{g_2}{f_2}$$
and the other three terms can be computed analogously.
\subsection{Higher derivatives}

Some of the thermodynamical identities involve higher derivatives and
we indicate briefly how to incorporate these into our scheme.
We use the self-explanatory notation $((a,b,c),d,e)$ for second
derivatives.  Thus $((3,1,2),2,1)$ is just $f_{12}$.  Note that this 
notation allows for such derivatives as 
$\left(\dfrac{\partial}{\partial T} \left(\dfrac{\partial E}{\partial p}
\right )_V\right )_S $ which is $((8,1,2),3,4)$.  Once again, we can express
all such derivatives (there are now 18,816 of them) in terms of  $x$, $y$, $f$,
$g$ and their partials (now up to the second order) using the chain rule.
For
$$((a,b,c),i,j)=((a,b,c),1,2)(1,i,j)+((a,b,c),2,1)(2,i,j) $$
and 
$(a,b,c)$, $(1,i,j)$ and $(2,i,j)$ can be dealt with using the
above tables.

\subsection{Derived quantities and thermodynamical identities}

The reason why there is a plethora of thermodynamical
identities is simple.  A large number of significant (and also insignificant)
quantities can be expressed or defined as simple algebraic combinations
of a very few (our primitive quantities $x$, $y$, $f$, $g$ and their partials).  Hence there are bound to be many relationships between them.
Our strategy to verify (or falsify) an identity is to use the above 
methods to express both sides 
in terms of  these quantities and check whether 
they agree.

Of course, there are myriads of such quantities and identities
and we can only bring a sample.  Thus we have
$$c_V =T \left(\frac{\partial S}{\partial T} \right )_V, $$
the heat capacity at constant volume, and
$$c_p = T
\left(\frac{\partial S}{\partial T} \right )_p, $$
the heat capacity at constant pressure. In our formalism,
$c_V=f (4,3,2)$ and $c_p=f(4,3,1)$, and so, from our tables,
$$c_V=f\dfrac{g_1}{f_1}, \quad c_p=f\dfrac{g_2}{f_2} .$$

Hence for the important quantities
$\gamma =\dfrac {c_p}{c_V}$ and ${c_p}-{c_V}$
we have  $\gamma =\dfrac {f_2 g_1}{f_1 g_2}$ and 
${c_p}-{c_V}=f\dfrac1{f_1 f_2}.$

Further examples are 
$$l_V = \left(\dfrac{\partial S}{\partial V} \right )_T=-(4,2,3), $$
the latent heat of volume increase, and
$$l_p = \left(\dfrac{\partial S}{\partial P} \right )_T=(4,1,3), $$
the latent heat of pressure increase.

Further definitions are:
$$m_V = \left(\frac{\partial S}{\partial V} \right )_p = (4,2,1) $$
and
$$m_p = \left(\frac{\partial S}{\partial p} \right )_V=-(4,1,2). $$

The coefficient of volume expansion at constant pressure is

$$\alpha_p = \frac 1 V \left(\frac{\partial V}{\partial T} \right )_p=
\frac 1 y (2,3,1) $$
 and the isothermal bulk modulus of elasticity is
$$B_T = -V 
\left(\frac{\partial P}{\partial V} \right )_T=-y(1,2,3).$$ 

Then $K_T = \dfrac 1{B_T}=\dfrac {-1} {y(1,2,3)}$ is the isothermal 
compressibility.

We illustrate our method  by verifying the simple identity:
$$c_p-c_V=T\left(\dfrac{\partial P}{\partial T} \right)_V
\left(\dfrac{\partial V}{\partial T} \right)_p .$$
Using the tables above, we can easily compute both sides in terms of our 
primitive expressions
and get $\dfrac f{f_1f_2}$ in each case. 
\subsection{Computing $(a,b,c)$ and $((a,b,c),d,e)$}
We can summarise these results in the following formulae:
$$[a,b;c,d]=\frac{(a , 1,2 )( b,2 ,1 )-( a,2 ,1 )(b ,1 ,2 )}{(c ,1 ,2 )(d ,2 ,1 )-(c,2 ,1 )(d ,1,2)}
$$
and so
$$(a,b,c) =[a,c;b,c]=\frac{(a , 1,2 )( c,2 ,1 )-( a,2 ,1 )(c ,1 ,2 )}{(b ,1 ,2 )(c ,2 ,1 )-(b,2 ,1 )(c ,2 , 1)},$$
which allow us to systematically compute any of the derivatives of the form 
$(a,b,c)$ in terms of our primitives $x$, $y$, $f$, $g$, $f_1$, $f_2$, $g_1$ 
and $g_2$, using the above data basis for expressions of the form $(a,1,2)$
and $(a,2,1)$.

For the second derivatives we substitute $$\phi=\dfrac{(a , 1,2 )( c,2 ,1 )-( a,2 ,1 )(c ,1 ,2 )}{(b ,1 ,2 )(c ,2 ,1 )-(b,2 ,1 )(c ,1,2)}$$
into the formula

$$(\phi,d,e)=\frac{(\phi , 1,2 )(d ,2 ,1 )-( \phi,2 ,1 )(d ,1 ,2 )}{(d,1 ,2 )(e ,2 ,1 )-(d,2 ,1 )(e ,1,2)}
$$
to compute $((a,b,c),d,e)$ in terms of our primitive terms (this time with the 
first and second derivatives of $f$ and $g$).
The advantage of these formulae is, of course, that one can write 
a simple programme to compute them.
(It is always tacitly assumed in the above formulae 
that the appropriate conditions which allow
a use of the inverse function theorem hold).

It remains only to produce the corresponding data basis for 
second derivatives, i.e. to express all of the non-trivial 
quantities of the form $((a,b,c),d,e)$
with $b$, $c$, $d$ and $e$ either $1$ or $2$ in terms of 
$x$, $y$ and $f$ and $g$ and their partials.
Of course, $((3,1,2),1,2)$, $((3,2,1),1,2)$, $((3,2,1),1,2)$, $((3,2,1),2,1)$
are just $f_{11}$, $f_{12}$ (twice) and $f_{22}$.  Similar identities hold for
the partials of $g$.

Further,
\begin{eqnarray*}
((5,1,2),1,2))&=&-g_1f_1-gf_{11};\\
((5,1,2),2,1))&=&1-g_2f_1-gf_{12};\\
((5,2,1),1,2))&=&-g_1f_2-gf_{12};\\
((5,2,1),2,1))&=&-g_2f_2-gf_{22}.\end{eqnarray*}

Note that the two expressions for the mixed partial coincide, since
$f_1g_2-f_2g_1=1$.

Similarly,
\begin{eqnarray*}
((6,1,2),1,2))&=&f_1g_1+fg_{11};\\
((6,1,2),2,1))&=&1+f_2g_1+fg_{12};\\
((6,2,1),1,2))&=&f_1g_2+fg_{12};\\
((6,2,1),2,1))&=&g_2f_2+fg_{22}.\end{eqnarray*}

\begin{eqnarray*}
((7,1,2),1,2))&=&-g_1f_1-gf_{11};\\
((7,1,2),2,1))&=&-g_2f_1-gf_{12};\\
((7,2,1),1,2))&=&-1-g_1f_2-gf_{12};\\
((7,2,1),2,1))&=&-g_2f_2-gf_{22}.\end{eqnarray*}

and
\begin{eqnarray*}((8,1,2),1,2))&=&g_1f_1-fg_{11};\\
((8,1,2),2,1))&=&g_1f_2+fg_{12};\\
((8,2,1),1,2))&=&-1+f_1g_2+fg_{12};\\
((8,2,1),2,1))&=&-g_2f_2+fg_{22}.\end{eqnarray*}

We emphasise that the numerical code for the  various thermodynamical
quantities is a mere construct to facilitate their computation
(ideally with the aid of suitable software) and that the final goal is to
express them all in terms of the basic quantities ($x$, $y$, $f$, $g$
and the partials of the latter).  It is then a routine matter to
translate these into the standard terminology of thermodynamics
if so required. 

\subsection{A notational survival kit}
In this treatment we have used three notations---the mathematically neutral
symbols $x$, $y$, $u$ and $v$ etc. (to develop the mathematical theory 
which is independent of any reference to thermodynamics), the standard 
thermodynamical
terminology $p$, $V$ etc. (for readers interested in the thermodynamical 
interpretation) and finally the numerical code (to systematise
the computations of derived quantities and our approach  to 
thermodynamical identities).  
For the convenience of the reader we give a dictionary of the relationships 
between them:
$$\begin{matrix}
1&2&3&4&5&6&7&8\\
p&V&T&S&G&H&F&E\\
x&y&u&v&E^{13}&E^{14}&E^{23}&E^{24}.
\end{matrix}$$
Of course, $p$ is pressure, $V$ volume, $T$ temperature, $S$ entropy
and $G$, $H$, $F$ and $E$ are free enthalpy, enthalpy, free energy
and energy respectively. 

\section{Existence and uniqueness of $S$-transversals}
In this section, we consider in more detail the restraints which are imposed 
on families of curves by the Samuelson area condition. We begin with a
special situation which we can compute directly and then show how to reduce the
general case to it. 
We show that,
given any family of level curves, there always exist  families 
of intersecting level curves which satisfy the area condition, 
and the method of proof allows us to compute these curves explicitly
in many interesting cases. Speaking loosely, this means that   
in thermodynamics to every family of isotherms 
(adiabats) there correspond  families of possible adiabats (isotherms) 
(this subject
is discussed more carefully below).  Since the area condition is 
not very demanding, there are infinitely many
collections of level curves which are $S$-transversal to a given family; 
but we show that if we are given two curves which
are transversal to the latter (in the differential geometric sense)
then they can be embedded in an essentially unique fashion into a
system of $S$-transversal curves.  This allows us, for example, to
write down all families which are $S$-transversal to  
the set of isotherms for the van der Waals gas.
We include some examples which we found to be of 
interest in a later section.

\subsection{A  special case}

 We begin by investigating the questions of existence and uniqueness 
when  one of the families consists of lines 
parallel to one of the axes, in this case, the $x$-axis.  For this example,
 elementary computation shows that, as claimed, we can always describe all
 other possible families of level curves which are $S$-transversal to the 
first one.  Moreover, in this case, it is also straightforward to show that 
knowledge of two curves determines the whole  family. 
 Remarkably, as we show in the next section, the general case can 
be reduced to this one, thus allowing a simple derivation of the basic theorems.
If the first family of curves is calibrated, i.e., 
they are the level curves of a 
particular potential function $u$, then any {\it single} curve suffices
to determine the second family.

So let the $v$-foliation consist of the
lines parallel to the $x$-axis (i.e., where
 $v(x,y) = y$), where in order to avoid topological difficulties we 
suppose that our potentials are defined on a product of intervals.  (We are 
exchanging the roles of $u$ and $v$ here and will 
compute all $u$-functions which are transversal to this $v$).

Then we know
that
if the  foliation induced by the potential $u$ is $S$-transversal, 
we can recalibrate $u$ and $v$ so that the Jacobian is
identically one.  If we assume that the $u$ foliation has already been
recalibrated and that the recalibration of $v$ is
$v(x,y) = c(y)$, then a straightforward computation  shows
that  $J = c'(y)u_x$ and so
$u$ must have the form:
$$u(x,y) = a(y) x + b(y)$$
where $b$ is an arbitrary smooth function of one variable and $a$
is such that $a(y) c'(y) = 1$.

We can state this formally as follows:

\begin{theor}  The function $u$ is $S_1$-transversal to the function 
$v$ = $c(y)$ if and only if $u$ has the form
$u(x,y) = a(y) x + b(y)$ where $a(y) = \dfrac 1 {c'(y)}$.
(Note that we are assuming here that $c$ is a diffeomorphism between two 
intervals of the line).
The function $u$ is $S$-transversal to the function $v = y$ if and only
if $u$ has the form
$$u(x,y) = \phi (a(y) x + b(y))$$
where  $\phi$ is a diffeomorphism between two intervals of the line,
and $a$ and $b$ are any two smooth functions of one variable, for which
$a$ has no zeros.
\end{theor}

Since it will often be convenient to switch the roles of
$u$ and $v$, or $x$ and $y$ respectively, in the above, we document 
the corresponding formulae:
$$u(x,y) = c(x), \quad v(x,y) = a(x) y + b(x).$$
We now consider  the question of uniqueness in the above situation.
By virtue of the
general theory developed in the next section, this will suffice to cover 
the general case.  
Our starting point is
the typical pair of $S_1$-transversal functions
$$u(x,y) =  a(y) x + b(y), \quad v(x,y) = c(y)$$
for arbitrary (generic) functions $a$ and $b$ (of one variable), with 
$a c' = 1$ (i.e., $c$ is a primitive of $\dfrac 1a$)
for the case where the $v$-lines are the parallels to the $x$-axis. 
We now suppose that we have two transversals to the $x$ axis which 
we want to incorporate into a family of geometric adiabatics.  We can 
suppose that the curves correspond to the values $u =  0$ and $u = 1$.
If $c = 0$, then we have
$$x = - \frac {b(y)}{a(y)}$$
and, for $c = 1$,
$$x =  \frac {1 - b(y)}{a(y)}.$$
We note now that if the $u$-level curves are to be $S_1$-transversal to the 
parallels to the $x$-axis, then they are transversal in the differential 
geometrical sense and so can be regarded as the  graphs of functions (more 
precisely, $x$ as a function of $y$).
Hence if we suppose that the \lq \lq adiabatics" $c = 0$, $c = 1$ have the form
$x = f_0(y)$, and $x = f_1(y)$, then a simple computation shows that
the general adiabatic has the form
$u = c$,
where $$u(x,y) = \frac x{f_1(y) - f_0(y)} - \frac {f_0(y)}{f_1(y) - f_0(y)}.$$

\subsection{A reduction} 

We now show that the general case can be reduced to the previous special case.
We begin with the remark that if we have any smooth
non-vanishing function  $f$ of two variables, say on the product of
 two intervals, then we can always
find a smooth vector field $(u,v)$ which has  $f$  as its Jacobi
function.

Probably the easiest way to do this, as was pointed out to us by
Michael Schm\"uckenschl\"ager, is to use a field of the form

$$u(x,y) = \phi(x,y),\quad  v(x,y) = \psi (y)$$
with $\phi$  a smooth
function of two variables, $\psi$ one of one
variable. (Such fields are called Knothe fields).
The above form also has 
the advantage  of leaving the level curves $y = d$ invariant.
The Jacobian of the above function is $\phi_x (x,y) \psi'(y)$ and we can,
of course, easily choose the two free functions in such a way that
this product gives $f$. For example, in the case where $f$ is the constant 
function $1$, then we can take for $\psi$ any smooth
function of one variable and then $\phi$ is determined up 
to a function of $y$ alone , i.e., has the form 
 $\phi(x,y) = \frac 1 { \psi'(y)} x + \chi (y)$ where $\chi$ is an 
arbitrary smooth function of one variable\footnote{When we include 
such a formula we are, of course, tacitly assuming that the operations
carried out on the generic functions involved are legitimate.  
In this case this means
explicitly that we are assuming that the derivative of $\psi$ never vanishes
, i.e., that $\psi$ is a diffeomorphism.  This type of situation will 
occur frequently in the following and since it would be tedious to 
state the explicit assumptions on the generic functions which 
arise, we will rely on the reader to fill in the details.}

Using this result, we can prove the following:

\begin{theor}

Suppose that we have a foliation of part of the plane by the level 
curves of a suitable function  $u(x,y)$ (with non-vanishing gradient) 
which is defined on a domain 
(i.e., an open, connected subset) $G$ in $\bf R^2$. Then we can linearise
$u$ locally by means of an area-preserving mapping.  More precisely,
for each point $(x_0,y_0)$ in $G$ we can find a neighbourhood  $\tilde G$
of the point in $G$ and   
a function $v(x,y)$ on $\tilde G$
which is such that the mapping $(x,y) \mapsto (u(x,y),v(x,y))$ is 
area-preserving and maps the lines $u = c$ onto  lines parallel to 
the $y$-axis. 
\end{theor} 

This is another result which is part of mathematical folklore.

In order to prove it, we start with a foliation consisting of the level curves
of a potential $u$ and a point $(x_0,y_0)$.  Since the gradient of $u$ 
never vanishes, we can find a function $v$ which is transversal to $u$ 
in a neighbourhood of this point and we introduce the new variables
$X = u(x,y), \quad Y = v(x,y)$.  By the inverse function theorem we can 
suppose that this can be solved to obtain $x$ and $y$ as smooth functions 
of $X$ and $Y$, say $x = a(X,Y), \quad y = b(X,Y)$.

We now introduce further new variables $\tilde X$ and $\tilde Y$ of the form
$$\tilde X = \phi (X),\quad  \tilde Y = \psi (X,Y)$$
for suitable smooth functions $\phi$ and $\psi$ of one and two variables
respectively.  Then elementary calculations show that the
Jacobian of $\tilde X$ and $\tilde Y$ with respect to the 
variables $x$ and $y$ (but expressed in terms of the variables $X$ and $Y$) is
$$\phi'(X)\psi_2(X,Y) J(X,Y)$$ 
where $J(X,Y)$ is the Jacobian of $(u,v)$ with respect to $x$ and $y$, 
expressed
as a function of $X$ and $Y$ {\it via} $a$ and $b$ , i.e.,
$J(X,Y) = \bar J(a(X,Y),b(X,Y))$ where 
$\bar J(x,y) = u_x(x,y)v_y(x,y) - u_y(x,y)v_x(x,y)$.
We can clearly arrange for 
this to be identically one by using the freedom in the choice of $\phi$ 
and $\psi$ and this completes the proof.

\subsection{The general situation}

Using these results, we can now extend the existence and uniqueness 
results from the special case in which one of the foliations 
is parallel to the axis to any given $u$-foliation.

Since the method is explicit we can also use it to find all 
possible $S$-transversal
systems for several interesting special types of $u$-curves.  The method
used is as follows:  Suppose that we can find an area-preserving mapping
which maps the $u$-curves onto the lines parallel the the $x$-axis. Then
we can transfer the above example to this situation.  We remark that
this is in a certain sense a rigorous justification for a ploy of
Maxwell's,  who  argued from this special situation (for reasons of 
simplicity), assuming that his conclusions then carried over to the
general case (cf. the passage: \lq \lq For the sake of the distinctness 
in the figure, I have supposed the substance to be partly in the liquid 
and partly in the gaseous state, so that the isothermal lines are horizontal, 
and easily distinguished from the adiabatic lines, which slope downward to 
the right.  The investigation, however, is quite independent of any such 
restriction as to the nature of the working substance'', [Ma], p. 155).
 
We state this formally as a theorem:

\begin{theor}
Suppose that we are given a foliation of the plane which we take to be
the level  curves of a suitable function $u$.  Then there exists (locally)
a family of curves (the level curves of a potential $v$) which are 
$S$-transversal
to the level curves of $u$.  Furthermore, given any two curves which are 
transversal to the level curves of $u$,  then there  exists a unique family
of $S$-transversal curves which include the given two.

\end{theor}

\subsection{Examples of the uniqueness and existence results}

We bring some explicit computations in connection with the
question of the existence and uniqueness of  $S$-transversals
to some simple cases.
\paragraph{The ideal gas:}
We begin with the case of the adiabatics for the ideal gas.  In this case
we use the new variables
$X=x y,\quad Y= \dfrac1{\gamma-1}\log\left (xy^\gamma \right)$
to reduce to the simple case of transversals to the parallels to the
coordinate axes.

Then we have that two functions $u$
and $v$ where $v$ is a recalibration of  $xy^\gamma$ are $S_1$-transversal
if and only if they have the form 
$$u(x,y)=a\left(\dfrac1{\gamma-1}(\ln x+\gamma\ln y)  \right )x y+
b\left (\left(\dfrac1{\gamma-1}(\ln x+\gamma\ln y)  \right )\right ),$$
$$v(x,y)=c\left(\dfrac1{\gamma-1}(\ln x+\gamma\ln y)  \right )$$
where $c$ is a primitive of $\dfrac 1a$.

Similarly, two functions $u$
and $v$ where $u$ is a recalibration of  $xy$ are $S_1$-transversal
if and only if they have the form 
$$u(x,y)=c(xy),\quad 
v(x,y)=a(xy)\left (\dfrac1{\gamma-1}\left (\ln x +\gamma \ln y\right )\right )+
b(x y)$$
where $c$ is again a primitive of $\dfrac 1a$.

From these formulae it is easy to give the general form of functions $u$
which are $S$-transversal to the adiabatics of the ideal gas resp.
functions $v$ which are $S$-transversal to its isotherms. 

At this point we bring a concrete example related to the ideal gas
which was constructed to answer a question of Samuelson.
Suppose that we are given $v(x,y) = x y$ and require an $S$-transversal 
function $u$ which interpolates between the curves $x y^2 = 1$ and $x y^3 = 10$
(i.e., two adiabatics corresponding to distinct cases of the ideal gas).  Then 
a simple computation shows that

$$u(x,y) = \frac {2\ln (xy^2)}{\ln (10 x y)}$$
is such that the contour $u = 0$ is the first curve, while $u = 1$
is the second one.  Interestingly, this then forces $u$ to contain
adiabatics of all the intermediary exponents, as the reader can easily verify.
(This is an example where the representations are only valid locally,
since any two curves of the form $xy^2=c$ and $xy^3=d $ will cross).

Analogous considerations lead to the following result:
\begin{theor}
Let $v$ be $S$-transversal to $u = xy$.  Then if two level curves of $v$
 have the form $xy^\gamma$ constant for a fixed $\gamma$, $v$ is a 
recalibration of $xy^\gamma$.
\end{theor}

\paragraph{The van der Waals gas:}
We now turn to the van der Waals equation. In order to simplify the
notation, we 
use the following solutions of the Maxwell relationships:

$$u=\left(x+\frac 1{y^2}\right)(y-1);$$

$$v=\dfrac 1{\gamma-1}\left (\ln\left(x+\frac 1{y^2}\right)+\gamma \ln(y-1)\right ).$$

 In this case
we use the new variables
$$X=\left(x+\frac 1{y^2}\right)(y-1)$$
and
$$Y=\dfrac 1{\gamma-1}\left (\ln\left(x+\frac 1{y^2}\right)+\gamma \ln(y-1)\right ) $$
to reduce to the simple case.

Then we have that  two functions $u$
and $v$ where $v$ is a recalibration of  $Y$ are $S_1$-transversal
if and only if they have the form 
$$u(x,y)=a(Y)X+
b(Y),\quad 
v(x,y)=c(Y)$$
where $c$ is a primitive of $\dfrac 1a$.

Similarly, two functions $u$
and $v$ where $u$ is a recalibration of  $xy$ are $S_1$-transversal
if and only if they have the form 
$$u(x,y)=c(X),\quad 
v(x,y)=a(X)Y+
b(X)$$
where $c$ is again a primitive of $\dfrac 1a$.
 
From these formulae it is again easy to give the general form of functions $u$
which are $S$-transversal to the adiabatics of the van der Waals gas, or
functions $v$ which are $S$-transversal to its isotherms. 

Similar methods can be applied to the Feynman gas which are now discussed. 
\section{Five basic models} 

We conclude by collecting some explicit computations for various gas models: 
beginning with the ideal gas.

\subsection{The ideal gas}  Here the recalibration is
$u=xy$, $v=\dfrac1{\gamma-1}(\ln xy^\gamma)$.  In this case we can explicitly
compute the relationships, which are obtained by
permuting the variables to get:

$x = e^{(\gamma-1)v} y^{-\gamma},\quad u = e^{(\gamma-1)v} y^{-\gamma+1}$;

$x = \dfrac u y, \quad v = \dfrac1{\gamma-1}\ln u +\ln y$;

$y = \dfrac u x, \quad v =\dfrac\gamma{\gamma-1}\ln u -\ln x $;

$y = e^{\frac{\gamma-1}\gamma v} x^{\frac{-1}{\gamma}},
\quad y = e^{{\frac{\gamma-1}\gamma}v}x^{\frac{\gamma-1}{\gamma}}$.
 
The reader can check that the four Maxwell relations are indeed valid and
thus compute the corresponding energy fuctions.  We shall shortly describe a
simpler and more systematic way of doing this.

\subsection{A generalisation of the ideal gas}
We now consider a simple generalisation of the ideal gas. This will 
presumably not describe any real gas (but see the remarks on Nernst's 
law below):
we include it since the
results 
are particularly
transparent.  The starting point is the equation
$$u = x^a y^b, v = x^c y^d.$$
with $a \neq 1$, $b \neq 1$ and $ab-cd \neq 0$.
\noindent The canonical recalibration
is
\begin{eqnarray*}U&=&\frac 1{\sqrt J}\left(\frac
    1{d-c-J+1}\right)x^{a(d-c-J+1)}y^{b(d-c-J+1)}\\
V&=&\frac 1{\sqrt J}\left(\frac
  1{a-b-J+1}\right)x^{c(a-b=J+1)}y^{d(a-b-J+1)}.\end{eqnarray*}
where $J=ad-bc$,
and so, for the special case $J=1$, (which we can always achieve by 
means of a simple recalibration),\\
\begin{eqnarray*}\frac 1{d-c}&x^{a(d-c)}y^{b(d-c)}\\
\frac 1{a-b}&x^{c(a-b)}y^{d(a-b)}\end{eqnarray*}

We have included this example since the results have a pleasing 
simplicity and symmetry (see, in particular, the further computations below).
The case of an ideal gas can be obtained by setting $a=c=1$ and  letting
$b$ tend to $1$ but there are some subtleties involved, as the presence
of the logarithmic term in the recalibration of the ideal gas would suggest.
\subsection{The van der Waals gas}
The natural calibrations are
$$u = \left (x+\dfrac a {y^2}\right )(y-b)\qquad v = \dfrac 1{\gamma-1}\ln
\left( 
\left (x+\dfrac a {y^2} \right )
(y-b)^\gamma \right ).$$

In this case the computations for 
computing $E^{23}$
and $E^{24}$ can be carried out by hand and we get:

$$x - \frac{u}{y-b}+\frac a{y^2},\quad v=\ln(y-b)+\frac 1{\gamma-1} \ln u,
$$
and 
$$x=e^{(\gamma-1)v}(y-b)^{-\gamma},\quad u=e^{(\gamma-1)v}(y-b)^{1-\gamma}.
$$
Again this can be used to compute the two energies, but see below
for a more systematic treatment which gives all such functions.
\subsection{The Feynman gas}
Here $u=x y$ and $v=x y^{\gamma(xy)}$ for a function $\gamma$
of one variable. This example was introduced by Feynman (see [Fe])
to cope with the fact that, in a real gas, the adiabatic index depends on 
temperature. This is a case where one genuinely requires the equation
in [Co1] to verify that it is a Samuelson configuration.
This turns out to be the case and the recalibrations (which are not computed by
Feynman)
are
$$U = \phi(x y) \qquad V=\ln(x y^{\gamma(x y)})$$
where $\phi$ is a primitive
of $\dfrac 1{\gamma-1}$.
This is the first example which we have met where we genuinely have
to recalibrate temperature (i.e., Boyle's law holds only in the weak form
that $pV$ is constant for constant temperature).  
For a discussion of the relevance of 
such recalibrations, see Chang [Ch].  The recalibrations introduced provide
an at least qualitative explanation of the diagram on p. 78 of this reference,
which displays comparative data of Le Duc on spirit thermometers.

Here we can solve for the cases where $u$ and $x$, or  $u$
and $v$, are the independent variables.
(We would like to thank P.F.X. M\" uller who pointed out this passage 
in Feynman's text to us).
\subsection{A synthesis} 
We can include all of the above (except the second example) in
the form:

$$u = \left(x+\frac a{y^2}\right)(y-b) \qquad 
v= \left(x+\frac a{y^2}\right)(v-b)^{\gamma(u(x,y))}.$$
with recalibrations
$$u = \phi\left( \left(x+\frac a{y^2}\right)(y-b)\right) \qquad 
v= \ln\left(\left(x+\frac a{y^2}\right)(v-b)^{\gamma(\phi^{-1}(u(x, y))}\right).$$
Once again, $\phi$ is a   primitive of $\dfrac1{\gamma-1}$.
This is another case where it seems hopeless to check 
that this represents a Samuelson configuration 
without the theory and computational methods developed here.

\subsection{A gallimaufry of formulae}
For completeness, we now bring a list of the expressions $(i,1,2)$ and $(i,2,1)$
for the substances introduced above.  We emphasise again that we include
these results since they allow us to compute the various energy 
functions and further thermodynamical quantities without explicitly calculating
the various permutations
of the variables implicit in the definitions.  We have found no indication
in the literature that this is possible.  Again we start with the ideal gas:
\paragraph
{The ideal gas:}
For the  ideal gas
\begin{eqnarray*}
(5,1,2) &=&y-\frac{y \log \left(x y^\gamma\right)}{\gamma-1};\\
(5,2,1) &=&-\frac{x \log \left(x y^\gamma\right)}{\gamma-1};\\
(6,1,2)&=&\frac{y \gamma}{\gamma-1};\\
(6,2,1)&=&\frac{\gamma x}{\gamma-1};\\
(7,1,2)&=&-\frac{\left(y \log\left(x y^\gamma\right)\right)}{( \gamma-1)};\\
(7,2,1) &=&-x - \frac{(x \log(x y^\gamma))}{(\gamma-1)};\\
(8,1,2)&=&\frac y {(\gamma-1)};\\
(8,2,1)&=&\frac{x}{(\gamma-1)}.
\end{eqnarray*}

\paragraph
{The generalisation of the ideal gas:}
Here
$$(5,1,2)= \dfrac {by} {b-a}, \qquad (5,2,1) = 
\dfrac{bx}{b-a},$$

$$ (6,1,2)=\dfrac{dy}{d-c},\qquad (6,2,1)=\dfrac{dx}{d-c},$$

$$(7,1,2)=\dfrac{ay}{b-a}, \qquad (7,2,1)=\dfrac{ax}{b-a},$$

$$(8,1,2)= \dfrac{cx}{d-c},\qquad (8,2,1)=\dfrac{cy}{d-c}.$$
Hence the energy functions are given by
 $E^{12}=\dfrac{bxy}{b-a}$, $E^{13}=\dfrac{dxy}{d-c}$, 
$E^{23}=\dfrac{axy}{b-a}$
and
$E^{24}=\dfrac{cxy}{d-c}$.

Note the pleasing symmetry of these results.  We have found them 
useful as a litmus test
for the validity of thermodynamical identities.

We remark here that although this model may not  correspond to any real gas,
it does have the advantage that it satifies Nernst's law---the third law
of thermodynamics in the precise form given in [La]---i.e., that
if we express the entropy $S$ as a  function of $p$ and $T$ 
or of $V$ and $T$, then we get the form  $S= P_1(p)T^n$ or $S= P_2(p)T^m$
for suitable positive indices $n$ and $m$ and functions $P_1$ and 
$P_2$ of pressure. 
We know of no other explicit model for a real gas
which has this property.

\paragraph
{A non-example}

Continuing on the theme of Nernst's law, we note that we have examined
the Feynman model in this respect (for natural choices of $\gamma$)
and found that it again failed to reproduce this phenomenon---the
problem lies in the logarithm term in the recalibration of entropy.
In view of the above remark, it was then tempting to
combine the Feynman model and the above generalisation of the ideal gas, i.e. 
to consider the case
$$u=x^a y^b,\quad v=x^c y^{\gamma(xy)}.$$
Unfortunately, these functions do not normally satisfy the $S$-condition.
Despite this disappointment, this computation at least shows the usefulness
of the P.D.E. characterisation of the latter condition, in particular
that it can be used to eliminate possible models which cannot be recalibrated
to satisfy the Maxwell relations. 

\paragraph
{The van der Waals gas:}
\begin{eqnarray*}
(5,1,2)&=&\dfrac{(-b + y)}{(\gamma-1)};\\
(5,2,1)&=&-\frac{\left(-\dfrac{2 (y-b) a}{y^3}+\dfrac{a}{y^2}+x\right) \log
   \left(\left(\dfrac{a}{y^2}+x\right) (y-b)^\gamma\right)}{\gamma-1};\\
(6,1,2)&=&y -\dfrac{ ((-b + y) \log\left((x + \dfrac a{y^2}) (-b + y)^\gamma\right )])}{(\gamma-1)};\\
(6,2,1)&=&-\frac{\left(-\dfrac{2 (y-b) a}{y^3}+\dfrac{a}{y^2}+x\right) \log
   \left(\left(\dfrac{a}{y^2}+x\right) (y-b)^\gamma\right)}{\gamma-1};\\
(7,1,2)&=&y-\frac{(y-b) \log \left(\left(\dfrac{a}{y^2}+x\right) (y-b)^\gamma\right)}{\gamma-1};\\
(7,2,1)&=&\frac{(y-b)^{1-\gamma} \left(\gamma \left(\dfrac{a}{y^2}+x\right) (y-b)^{\gamma-1}-\dfrac{2 a (y-b)^\gamma}{y^3}\right)}{\gamma-1};\\
(8,1,2)&=&\frac{y-b}{\gamma-1};\\
(8,2,1)&=&\frac{(y-b)^{1-\gamma} \left(\gamma \left(\dfrac{a}{y^2}+x\right) (y-b)^{\gamma-1}-\frac{2
   a (y-b)^\gamma}{y^3}\right)}{\gamma-1}-x.\\
\end{eqnarray*}
\paragraph
{The Feynman gas:}
\begin{eqnarray*}
(5,1,2)&=&y-y \log \left(x y^{\gamma(x y)}\right) \phi'(x y);\\
(5,2,1)&=&-x \log \left(x y^{\gamma(x y)}\right) \phi'(x y);\\
(6,1,2)&=&\frac{\phi(x y) \left(y^{\gamma(x y)}+x \log (y) \gamma'(x y) y^{\gamma(x
   y)+1}\right) y^{-\gamma(x y)}}{x}+y;\\
(6,2,1)&=&{\phi}(x y) \left(\frac{\gamma(x y)}{y}+x \log (y) \gamma'(x y)\right);\\
(7,1,2)&=&-y \log \left(x y^{\gamma(x y)}\right) {\phi}'(x y);\\
(7,2,1)&=&-\log \left(x y^{\gamma(x y)}\right) {\phi}'(x y) x-x;\\
(8,2,1)&=&\frac{y^{-\gamma(x y)} {\phi}(x y) \left(y^{\gamma(x y)}+x \log (y) \gamma'(x y)
   y^{\gamma(x y)+1}\right)}{x};\\
(8,2,1)&=&{\phi}(x y) \left(\frac{\gamma(x y)}{y}+x \log (y) \gamma'(x y)\right)-x.
\end{eqnarray*}
In reading Feynman's treatment, one gains the impression that he is tacitly
assuming that the formulae for his model are obtained simply by plugging
a variable $\gamma$ into those for the ideal gas.  The presence of terms
involving the derivative of $\gamma$ in the above show that this is not 
the case
(for example, in the formulae for the important quantities  $c_p$, $c_V$
and their difference and quotient).

It is an easy task to compute the above quantities for the combined Feynman
and van der Waals gas (using Mathematica), but the results are too elaborate to 
be included here.

\section{Final remarks}

The mathematics of thermodynamics have never ceased to fascinate mathematicians,
who generally experience a sense of unease at the standard representations,
in particular of the laws of thermodynamics as an axiom system
(see, for example, [Se] for a critical evaluation). 
There have been many attempts to put them on a 
solid basis.  We mention,
in particular, Caratheodory [Ca1] and [Ca2], Lieb and Yngvason [Li]
and Truesdell [Tr].  We  have, of course, been influenced by these treatments
and, inevitably, there are certain common points.  However, we believe that
our approach is sufficiently original to justify its presentation.
Thus in [Li] the ordering \lq \lq adiabatically accessible from'' is centre 
stage but apart from that the method is completely different.
We know of two systematic approaches to thermodynamical identities
(Bridgman [Br] and Jayne [Ja]) 
and they have influenced our treatment. 
Thus the idea of 
using Jacobians to  derive identities can be found in the latter\footnote{In Jayne's notation $[A,B]$ 
corresponds to our
$[A,B;x,y]$ 
for an unspecified pair of quantities $x$ and $y$.  The latter
can be freely chosen for any specific computation and so are not explicitly
documented in his symbolism.  Regardless of this choice, we always have 
that our $[A,B;C,D]$ is his
$[A,B]/[C,D]$,
which establishes the relationship between our notation and that of Jayne.}.

In conclusion, we would like t o express our gratitude to Iain Fraser and 
Elena Kartashova, who read and commented on an earlier version of 
our manuscript.

\end{document}